\documentclass[%
 reprint,
nofootinbib,
 amsmath,amssymb,
 aps,
 normalem
]{revtex4-2}

\usepackage{graphicx}
\usepackage{dcolumn}
\usepackage{bm}
\usepackage{ulem}

\usepackage[linktocpage=true]{hyperref}

\renewcommand{\eqref}[1]{Eq. (\ref{#1})}

\usepackage{xcolor}

\newcommand{\mpl}{m_{\mathrm{Pl}}}

\newcommand{\gagcrit}{g_{a\gamma}^\mathrm{crit}}

\begin{document}

{\hfill KCL-PH-TH/2024-14 \vspace{-20pt}}

\title{Multimessenger signals from compact axion star mergers} 

\author{Liina Chung-Jukko$^{a}$}
\email{liina.jukko@kcl.ac.uk}
\author{Eugene A. Lim$^{a}$}
\author{David J. E. Marsh$^{a}$}

\vspace{1cm}
\affiliation{${}^a$Theoretical Particle Physics and Cosmology Group, Physics Department, King's College London, Strand, London WC2R 2LS, United Kingdom}

\begin{abstract}

Axion dark matter can form stable, self-gravitating, and coherent configurations known as axion stars,  which are rendered unstable above a critical mass by the Chern-Simons coupling to electromagnetism. We study, using numerical relativity, the merger and subsequent decay of compact axion stars. We show that two sub-critical stars can merge, and form a more massive, excited and critical star, which survives for a finite period before rapidly decaying via electromagnetic radiation. We find a rich multimessenger signal, composed of gravitational waves, electromagnetic radiation, and axion radiation. The gravitational wave signal is broken into two parts: a weak and broad signal from the merger, followed by a much stronger signal of almost fixed frequency from the decay. The electromagnetic radiation follows only the gravitational waves from the decay, while the axion signal is continuous throughout the process. We briefly discuss the detectability of such a signal. 

\end{abstract}

\maketitle



\section{Introduction} \label{sec:intro}

Axion stars are self-gravitating compact objects formed of the QCD axion ~\cite{Peccei:1977hh,weinberg1978,wilczek1978,Kim:1979if,Shifman:1979if,Dine:1981rt,Zhitnitsky:1980tq,Preskill:1982cy,Abbott:1982af,Dine:1982ah,DiLuzio:2020wdo} or axion-like particles ~\cite{Turner:1983he,Witten:1984dg,Khlopov:1985jw,Svrcek:2006yi,Arias:2012az,Marsh:2015xka}. If the axion forms part or all of dark matter (DM), these stars would be expected to be abundant inside DM halos (for initial formation mechanisms, see Refs.~\cite{Schive:2014dra,Levkov:2018kau,Widdicombe:2018oeo,Eggemeier:2019jsu,Chen:2020cef}). As a time-periodic solution of the Einstein-Klein-Gordon equations, axion stars belong to the solitonic class of oscillatons~\cite{Seidel:1991zh,Seidel:1993zk}.  In the regime of strong gravity, these stars are compact, relativistic objects, where the mass of the compact star goes as $M_s\sim \mpl^2/m$. Throughout this work, units of the axion mass, $m$, are absorbed by choice of length and time units, and we furthermore use geometrized units with $G=c=1$~\footnote{To convert from our simulation units, as displayed on the figures, to physical units, the unit of $m$ has to be multiplied by $\mpl^{-2}$ before doing the usual conversion from geometrized units with factors of $G$ and $c$.}. 

Axions are coupled to electromagnetism (EM) through a Chern-Simons term with coupling constant $g_{a\gamma}$. In Ref.~\cite{Chung-Jukko:2023cow} we studied the instability of compact stars induced by $g_{a\gamma}$. Compact axion stars decay into a less massive and less compact remnant for masses $M_s$ larger than a critical value, which is set by the coupling constant, $\gagcrit \propto M_s^{-1.35}$. We demonstrated how this decay process is due to parametric resonance between the axion star and the EM field, which causes the star to lose energy by emitting EM radiation at its characteristic frequency. More specifically, when the characteristic frequency of the oscillating axion star lies within the parametric resonance band of the underlying scalar field, $\phi$, the oscillations of the star drive photons out of the electromagnetic field, causing a strong burst of EM radiation. We discuss the shutting down of the parametric resonance process in more detail in Appendix \ref{sec:resshut}. The instability of compact axion stars in the strong gravity regime studied in Ref.~\cite{Chung-Jukko:2023cow} is in agreement with previous studies of the instability of non-compact axion stars in the weak field regime~\cite{Hertzberg:2018zte,Levkov:2020txo}, and other related work (e.g. Refs.~\cite{Preskill:1982cy,Abbott:1982af,Kephart:1986vc,Boskovic:2018lkj,Ikeda:2018nhb}).

\begin{figure*}
    \centering
    \includegraphics[width=19.0cm]{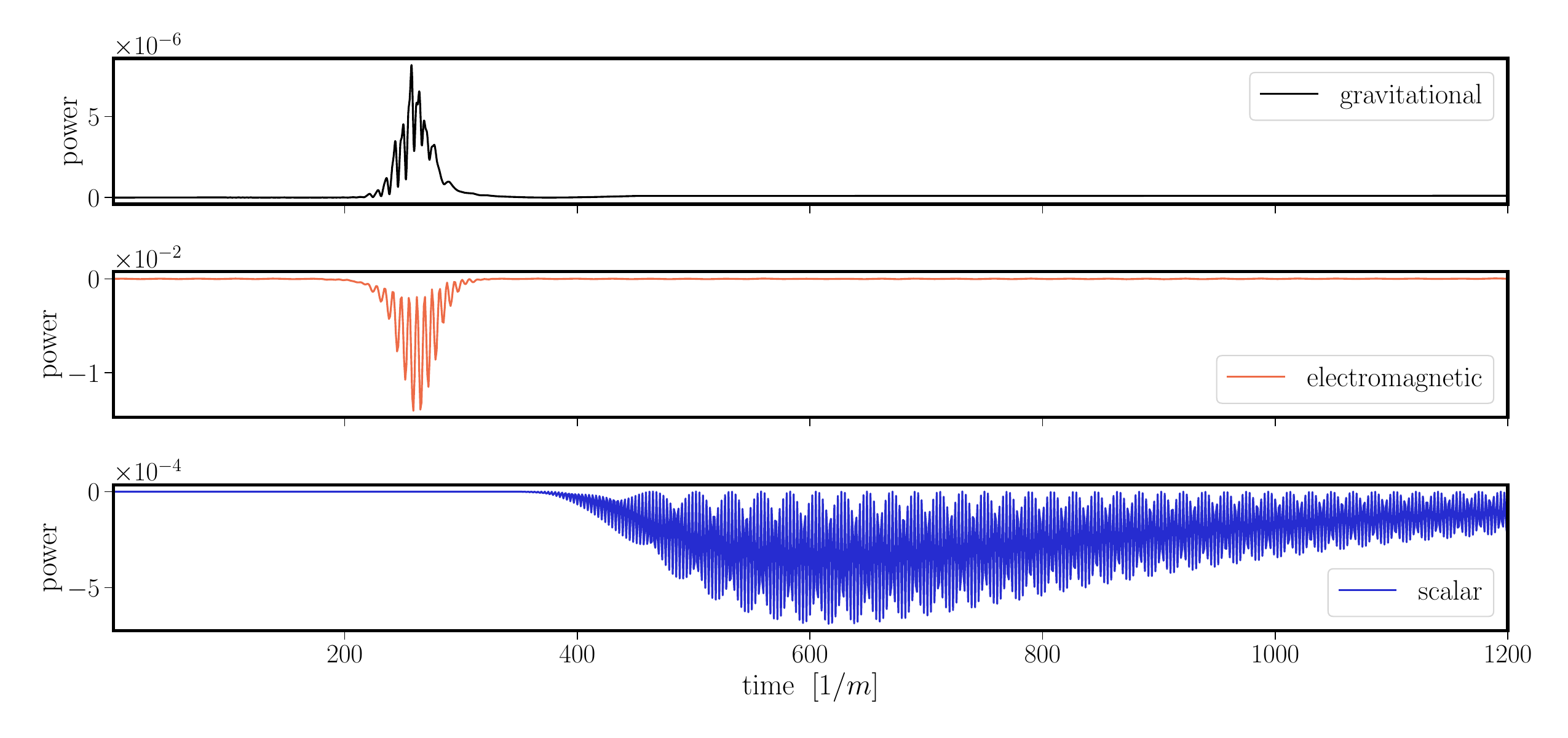}
        \vspace{-0.5cm}
    \caption{Multimessenger signal of axion star decay by electromagnetic instability, for an isolated super-critical star: gravitational power (luminosity) (\textit{top}), total flux of electromagnetic energy (\textit{middle}), and corresponding total flux of the scalar energy (\textit{bottom}).  Note that power is dimensionless in geometrized units. Signals are extracted over a sphere of radius $r=80m^{-1}$, and the power in electromagnetic and scalar radiation is obtained by integrating the corresponding flux. We note that the electromagnetic and GW signal reach the observer at the same time, as both travel at the speed of light, whereas the scalar emission is delayed as it is massive. The different frequency wave modes in the scalar emission are responsible for the long tail in the total scalar flux, as lower frequency modes travel more slowly away from the decaying star. The EM and scalar fluxes are calculated according to the method from Eq. 11 in Ref.~\cite{Clough:2021qlv}, and the gravitational power with the usual gravitational luminosity formula (see Appendix~\ref{eq:GWlum}).}
\label{fig:multimessenger_signal_single}
    \vspace{-0.5cm}
\end{figure*}

Hence, the existence of compact axion stars is restricted by their coupling to electromagnetism, which naturally raises the possibility of mergers as a formation mechanism for compact axion stars, as any ``pristine'' super-critical stars created in the early universe e.g. following Ref.~\cite{Widdicombe:2018oeo}, would rapidly decay. Therefore in the following we investigate the EM interaction of an axion star created by a head-on collision of two less compact but gravitationally stable stars. The original stars belong to the sub-critical compactness~\footnote{In this paper, we use the same definition for the compactness parameter $C$ as in \cite{Helfer:2018vtq}. For an axion star mass $M_s$, $C = GM_s / R$, where $R$ is the effective radius of the star (includes $95\%$ of its mass).} region found in Ref.~\cite{Helfer:2018vtq} where the merger leads to the formation of another excited but stable star (for more compact stars the merger leads to the prompt formation of a black hole). Compact axion stars are gravitational objects, and thus one expects their merger~\cite{Giudice:2016zpa} and subsequent decay to produce gravitational waves, in addition to the EM and axion (scalar) radiation seen already in Ref.~\cite{Chung-Jukko:2023cow}. We extract, for the first time, the full multimessenger signal in GW, EM and axions caused by axion star collisions and decays. An example signal extracted from the clean decay of an isolated compact axion star~\cite{Chung-Jukko:2023cow} is shown in Fig.~\ref{fig:multimessenger_signal_single} (the more complex signal from mergers will be shown later in the paper).

Our calculations are performed in the strong gravity regime, using the 3+1 numerical relativity code \textsc{GRChombo} ~\cite{Clough:2015sqa,Andrade:2021rbd,Radia:2021smk}, hence capturing all backreactions and other gravitational effects. We show that two stars originally in the stable region of the plane given by the two dimensionful parameters $(M_s, g_{a\gamma})$ (i.e. below the critical line $\gagcrit(M_s)$) can merge and form a highly excited, more compact star that correspondingly has a lower critical coupling, and is thus unstable electromagnetically.
This merged star then decays through parametric resonance just as the single stars studied in \cite{Chung-Jukko:2023cow}, emitting a strong EM burst.~\footnote{As in Ref.~\cite{Chung-Jukko:2023cow}, an EM seed is required to trigger the resonance: in our simulations such a seed is given by a weak propagating EM wave, modelling ambient radiation in the galaxy.} The GW signal of the decay in Fig.~\ref{fig:multimessenger_signal_single} does not exhibit any ringdown hierarchy as no black hole is formed. Instead, its frequency matches the axion mass, implying that it is primarily sourced by the quadrupole moment during the decay process. The GW signal from the merger is largely unaffected by the EM coupling, and agrees with the results of Ref.~\cite{Helfer:2018vtq} (which included axion self-interactions but no EM coupling). Thus, the multimessenger signal of compact axion star mergers has a distinct portrait: an initial weak GW signal from the merger, followed by a louder GW burst and accompanying EM signal, and finally a delayed and prolonged axion signal.

This paper is organised as follows: Section \ref{sec:theory} introduces our oscillaton model and the initial set up of our simulations. Section \ref{sec:mergerstory} outlines the story of two axion star colliding head-on, merging, and the merged star decaying through the electromagnetic instability demonstrated in \cite{Chung-Jukko:2023cow}. Section \ref{sec:GW} shows the gravitational wave signal from the electromagnetic decay of axion stars, and the combined signal from the merger and decay process, followed by the full multimessenger signal in Section \ref{sec:multimess}. Finally, we discuss the observational consequences of our findings in Section \ref{sec:discuss}.

\section{Theory} \label{sec:theory}

Our work in this paper uses the same oscillaton model as in our previous study \cite{Chung-Jukko:2023cow}. The total action is
\begin{equation} \label{eq: action}
\begin{aligned}
S=\int d^{4} x\sqrt{-g}&\left[ \frac{\mpl^2}{16\pi}R-\frac{1}{2} \partial_{\mu} \phi \partial^{\mu} \phi-\frac{1}{2} m^{2} \phi^{2}\right. \\
 &\qquad\left. -\frac{1}{4} F_{\mu \nu} F^{\mu \nu} -\frac{g_{a \gamma}}{4} \phi F_{\mu \nu} \tilde{F}^{\mu \nu} \right],
\end{aligned}
\end{equation}
where $\phi$ denotes the axion field, and $R$ the Ricci scalar. The electromagnetic field strength tensor and its dual are defined as
\begin{equation} \label{eq: fieldstrength}
F_{\mu \nu}=\partial_{\mu} A_{\nu} - \partial_{\nu} A_{\mu}, \quad \widetilde{F}^{\mu \nu}=\frac{1}{2 \sqrt{-g}} \varepsilon^{\mu \nu \rho \sigma} F_{\rho \sigma},
\end{equation}
with the totally antisymmetric Levi-Civita symbol $\varepsilon^{\mu \nu \rho \sigma}$, defined through $\varepsilon^{0123}=+1$.

The axion pseudoscalar field is coupled to the field strength tensor through the Chern-Simons term (last term in Eq.~\ref{eq: action}), where the coupling constant $g_{a\gamma}$ sets the strength of the interaction. Moreover, the stress-energy tensor can be calculated from the action in \eqref{eq: action} to solve Einstein's equations $G_{\mu \nu}=8 \pi \mpl^{-2} T_{\mu \nu}$ (see e.g. Ref.~\cite{Gorbar:2021rlt}), where we note that the Chern-Simons topological term does not enter the stress-energy tensor. 

The parametric resonance process, as described in Section \ref{sec:intro}, is driven by the sourced Maxwell's equation for the evolution of the field strength tensor, \eqref{eq: FEOM}, 
\begin{align}
\nabla^\mu \nabla_\mu \phi-m^{2} \phi&=\frac{g_{a \gamma}}{4} F_{\mu \nu} \tilde{F}^{\mu \nu}, \label{eq: PhiEOM} \\
\nabla_{\mu} F ^{\mu \nu}&=-g_{a \gamma} J^{\nu},\label{eq: FEOM}
\end{align}
where the current $J^{\nu}$ is given by $J^{\nu} = \partial_{\mu} \phi \tilde{F}^{\mu \nu}$. In general, resonance can happen provided that the photon frequency lies within the parametric resonance band. As we discussed in \cite{Chung-Jukko:2023cow}, the axion mass implies that the oscillations of the star go as $\omega \sim m \sim R_s^{-1}$, hence setting a bandwidth for the photon frequencies that can drive the parametric resonance process.

Our system of coupled axion star, electromagnetism and the metric is solved using \textsc{GRChombo}~\cite{Andrade:2021rbd, Radia:2021smk, Clough:2015sqa}. We follow the same procedure to find initial conditions for the axion star and the electromagnetic field as in our previous work \cite{Chung-Jukko:2023cow}. The axion star initial set up is built following Refs.~\cite{Seidel:1991zh, Helfer:2018vtq, Helfer:2016ljl, Alcubierre:2003sx, Michel:2018nzt,Seidel:1991zh, Urena-Lopez:2002ptf, Urena-Lopez:2001zjo}, while the implementation of the electromagnetic field follows the methodology introduced in Refs.~\cite{Zilhao:2015tya,Helfer:2018qgv,Gundlach:2005eh,Palenzuela:2009hx,Hilditch:2013sba}. Specifically, for the superposition of two axion stars as initial conditions, we use the profiles introduced in Ref.~\cite{Helfer:2018vtq,Helfer:2021brt, Evstafyeva:2022bpr} where the spatial metric is renormalized to reduced gauge artifacts. The EM initial conditions were set up as outlined in Ref.~\cite{Chung-Jukko:2023cow}, with the initial EM amplitude kept the same for all stars ($C_x = 0.001m_{\mathrm{Pl}}$), but a frequency of $k^{(x)} \equiv 2\pi/\lambda \sim 0.10m$ for the single, more massive star and $k^{(x)} \sim 0.12m$ for the colliding stars. 

As we solve the fully non-linear evolution equations numerically, our simulations include all backreactions. We implement periodic boundary conditions in the propagation direction of our electromagnetic plane wave seed to keep the radiation in the simulation box, and Sommerfeld boundary conditions in the remaining two directions. The Sommerfeld boundary conditions absorb outgoing radiation at the boundary of our simulation box which improves the accuracy of  gravitational wave signal extraction from the simulation. The evolution of the scalar field is not affected by the choice of boundary conditions, as the main effect of implementing Sommerfeld boundary conditions is to absorb outgoing gravitational and electromagnetic radiation. The convergence of our simulation, along with the convergence of the extracted gravitational waves, is discussed in Appendix \ref{sec:convergencetesting}.

\section{Mergers and electromagnetic instability} \label{sec:mergerstory}

\begin{figure*}
    \centering
    \includegraphics[width=17.0cm]{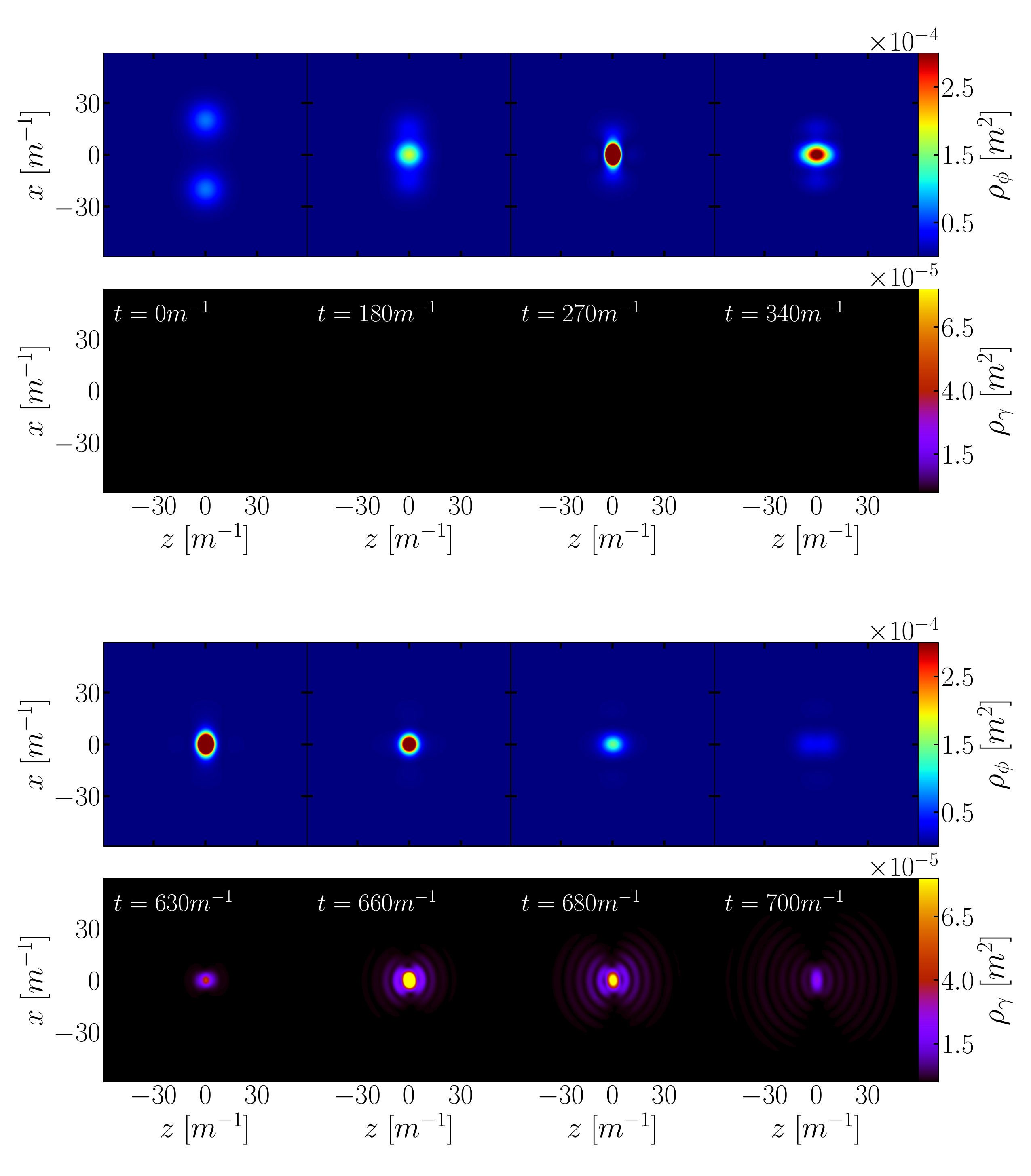}
        \vspace{-0.5cm}
    \caption{Evolution of the energy density in the scalar (\textit{blue panels}) and EM field (\textit{black panels}) as a slice through the center of our simulation box. We start with two stable stars of mass $M_s = 0.36 m_{\mathrm{Pl}}^2/m$ at $t=0m^{-1}$, which merge to form a highly excited, but stable star (around $t \sim 180m^{-1}$). Our coupling value is $g_{a\gamma} = 20 m_\mathrm{Pl}^{-1}$, implying that the final merger star is super-critical according to Ref.~\cite{Chung-Jukko:2023cow}, resulting in the decay of the merged star through parametric resonance (see the lower set of panels). This happens because, due to the higher mass, the critical coupling of the merged star is lower than the critical coupling of the original stars, allowing the parametric resonance process after the merger. The initial seed electromagnetic field (not visible on this scale) is polarised in the $x$-direction and propagating from the right to the left. A movie of our simulations can be found through \href{https://www.youtube.com/watch?v=SItsbjsr_tg}{this link}.}
\label{fig:panel}
    \vspace{-0.5cm}
\end{figure*}

Our story of two axion stars colliding head-on and merging, is captured in Fig.~\ref{fig:panel}, which demonstrates the evolution of the energy density in the scalar and EM fields over time in a slice through our simulation box. We start with two gravitationally and electromagnetically stable stars that collide and form a highly excited merged star that is stable gravitationally, but which after some time decays electromagnetically through parametric resonance. From now on, we will call the star resulting from the collision and merger of the two original axion stars ``merged star'', as opposed to the ``single star'' which is directly set up by the initial conditions.

More specifically, we start with compact axion stars in the sub-critical region demonstrated in \cite{Helfer:2018vtq}, where the head-on merger of these stars leads to a highly excited, more compact star. Our stars are initially separated by a distance of $40 m^{-1}$, and have a mass of $M_s = 0.36 m_{\mathrm{Pl}}^2/m$. Using the scaling relation between the critical coupling of a compact star and its mass $\gagcrit \propto M_s^{-1.35}$, which we demonstrated in Ref.~\cite{Chung-Jukko:2023cow}, this gives our initial stars a critical coupling of $\gagcrit \sim 24.6 m_\mathrm{Pl}^{-1}$.

Hence, we collide these stars with a coupling value of $g_{a\gamma} = 20 m_\mathrm{Pl}^{-1}$, which ensures that the interaction between the stars and the electromagnetic field is negligible before the merger, since the coupling value lies in the stable region of the $(M_s, g_{a\gamma})$ plane (under the critical line $\gagcrit(M_s)$), as demonstrated in Ref.~\cite{Chung-Jukko:2023cow}. In other words, the decay of the original stars through the parametric resonance process, as outlined in Section \ref{sec:intro}, is forbidden.

However, the stars collide and merge to form a single highly excited star, as shown in \cite{Helfer:2018vtq}. This merged star potentially possesses a critical coupling below the coupling set in the simulation ($g_{a\gamma} = 20 m_\mathrm{Pl}^{-1}$). We note that evaluating the critical coupling of this merged star analytically is not straightforward, as the star is highly excited with some ``breathing mode'' oscillation patterns and gravitational cooling by axion emission~\cite{Seidel:1993zk}, and hence does not map directly to the simple compact stars with critical line found in Ref.~\cite{Chung-Jukko:2023cow}. Nonetheless, approximating the merged star's effective radius by the location where the scalar energy density has dropped to $5\%$ of its central value, we compute the ADM mass of the merged star, which gives us an estimate of its critical coupling from the critical line in \cite{Chung-Jukko:2023cow}. We find $\gagcrit \sim 16 m_\mathrm{Pl}^{-1}$, which is below the simulated value. Hence, we expect parametric resonance to kick in after the merger and cause the merged star to decay. 

\begin{figure}[h!]
    \centering
    \includegraphics[trim=0.8cm 1.0cm 0.8cm 0.9cm, clip, width=9.0cm]{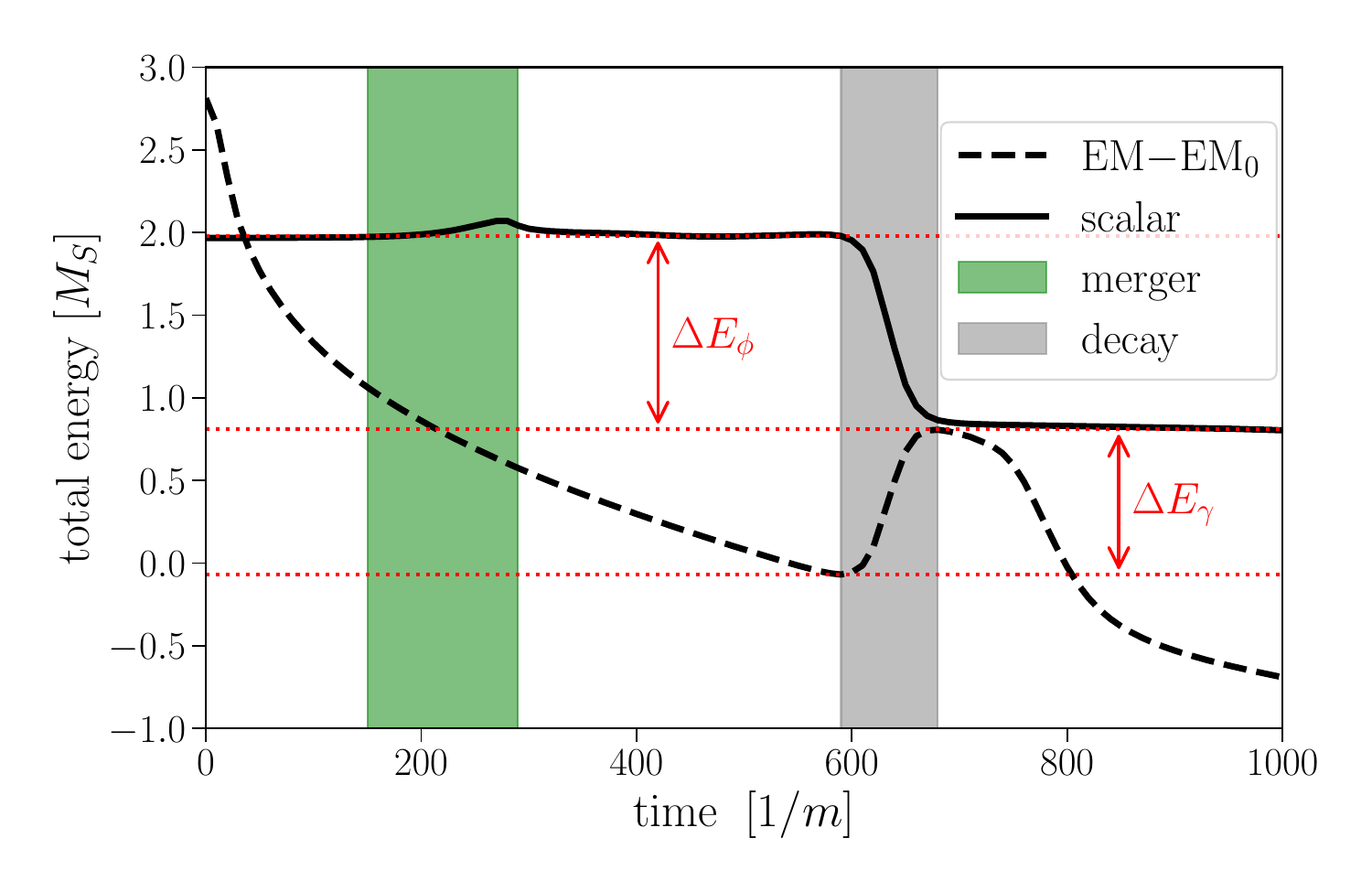}
    \caption{Total simulation box energy in the scalar (\textit{solid line}) and electromagnetic field (\textit{dashed line}) for the merger and electromagnetic decay process. The merger timescale is shown in green, and the decay through parametric resonance of the resulting highly excited merged star in grey. The energy is in units of the original colliding stars' mass. The electromagnetic energy $\mathrm{EM}_0$ immediately before the decay has been subtracted from the total; the decrease in the EM energy is due to the absorbing boundary conditions. Strong amplification of the EM field during the decay of the excited merged star is clearly visible ($\Delta E_{\gamma}$). The unphysical bump in the total scalar energy corresponds to the lapse being driven to its minimum value due to the large curvatures present at merger.}
    
    \label{fig: collenergy}
        \vspace{-0.5cm}
\end{figure}

This is indeed what we observe in our simulations, with the merging of the original stars shown in the upper set of panels of Fig.~\ref{fig:panel}, corresponding to $t=0m^{-1}$ to $t=340m^{-1}$, and the parametric decay of the merged star at later times $t=630m^{-1}$ to $t=700m^{-1}$ in the lower set of panels. The electromagnetic decay leaves behind a less compact remnant, and we discuss the corresponding shutting down of the resonance process in Appendix \ref{sec:resshut}.

The same merger and decay process is shown in terms of the total energy in the scalar and electromagnetic field in Fig.~\ref{fig: collenergy}. The green area corresponds to the merging process, while the grey area shows the time span of the parametric resonance decay of the merged star. We have scaled the energies to the mass of the original star ($M_s$), and hence the solid line representing the scalar energy starts at $2M_s$. The total scalar energy in the simulation box stays roughly constant before and after the merger of the stars, but decays through the parametric resonance process after $t \sim 600 m^{-1}$, as expected. Accordingly, the electromagnetic energy in the box is strongly amplified during parametric resonance. The decay process pictured in Fig.~\ref{fig: collenergy} is in perfect agreement with the decay of single compact stars that we studied in \cite{Chung-Jukko:2023cow}. 

We use Sommerfeld boundary conditions for our simulation box and hence the electromagnetic radiation is constantly being absorbed at the boundaries of our box, which explains its decrease until the parametric resonance kicks in. Furthermore, note the \emph{total} EM energy of the box is larger than that of the scalar -- this is due to the large box size we used in our simulations since the EM seed wave fills the entire domain. The key physical effect is the EM energy ``bump'' occuring at $t=600 m^{-1}$, i.e. $\Delta E_{\gamma} \sim \Delta E_{\mathrm{\phi}}$ where scalar energy is converted into EM energy via parameteric resonance.

\begin{figure}[t!]
    \centering
    \includegraphics[width=9cm]{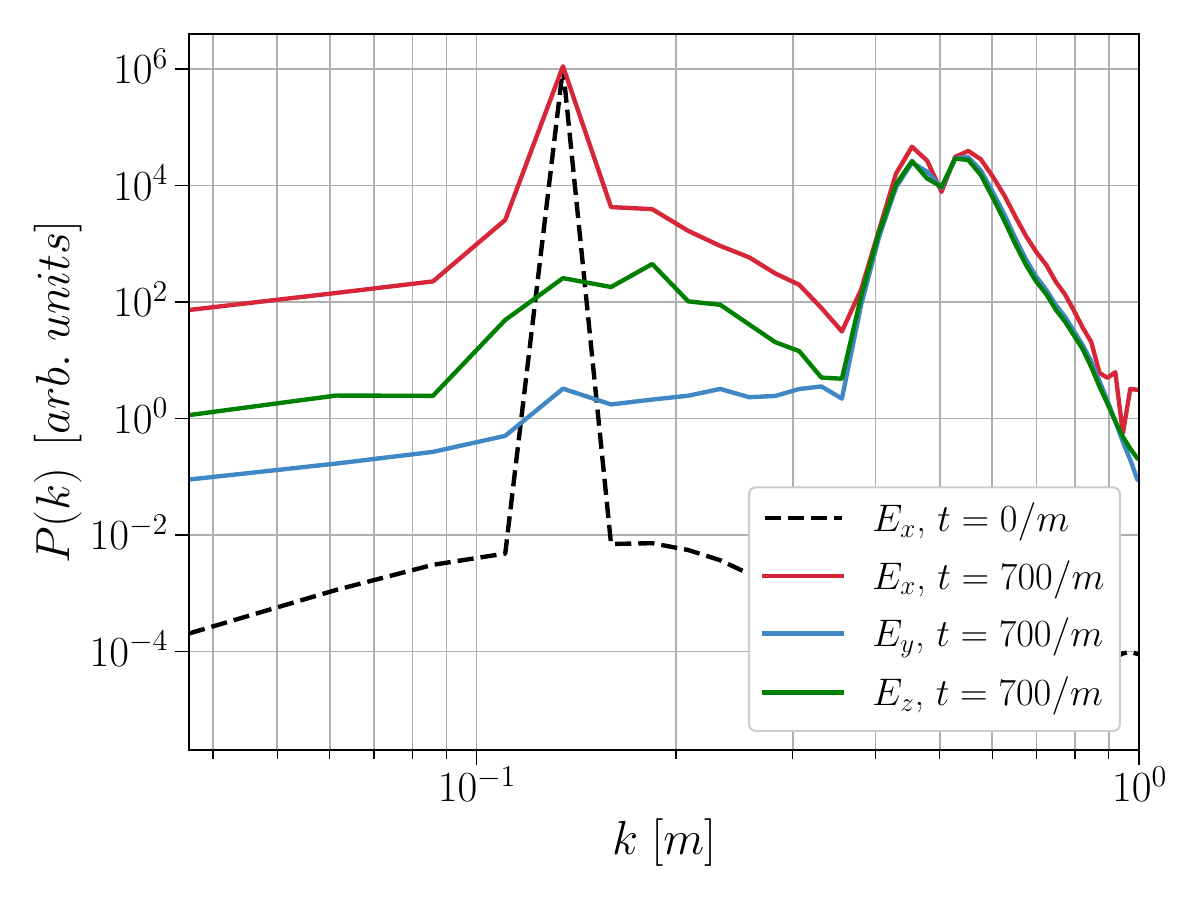}
        \vspace{-0.7cm}
    \caption{The power spectrum of the electromagnetic field at the start of the simulation ($t=0m^{-1}$) and after the decay of the merged star ($t=700m^{-1}$). A clear excitation in all the EM components around the characteristic frequency of the star, $k \sim 0.5m$, is visible. The $E_y$ component of the emission does not acquire energy at the seed frequency due to our Sommerfeld boundary conditions.}
    \label{fig: powerspec}
        \vspace{-0.5cm}
\end{figure}

The power spectrum of the electromagnetic field at the start of the simulation and right after the decay of the merged star has happened is shown in Fig.~\ref{fig: powerspec}. The dashed black line demonstrates the original seed (polarised in the x-direction), while the coloured lines corresponding to $t=700m^{-1}$ show all the components of the EM field ($E_x$, $E_y$, $E_z$), with a clear excitation around $k\sim0.5m$, which corresponds to the characteristic frequency of the star, and demonstrates parametric resonance. Similar to the power of a single star decay discussed in \cite{Chung-Jukko:2023cow}, the EM power is approximately isotropic. Interestingly, there is a small double peak around $k \sim 0.5m$, which is likely due to a slight difference in excitation frequency of the oscillations around the collision axis and that perpendicular to it.

\section{Gravitational waves} \label{sec:GW}

\begin{figure}[t!]
    \centering
    \includegraphics[width=9cm]{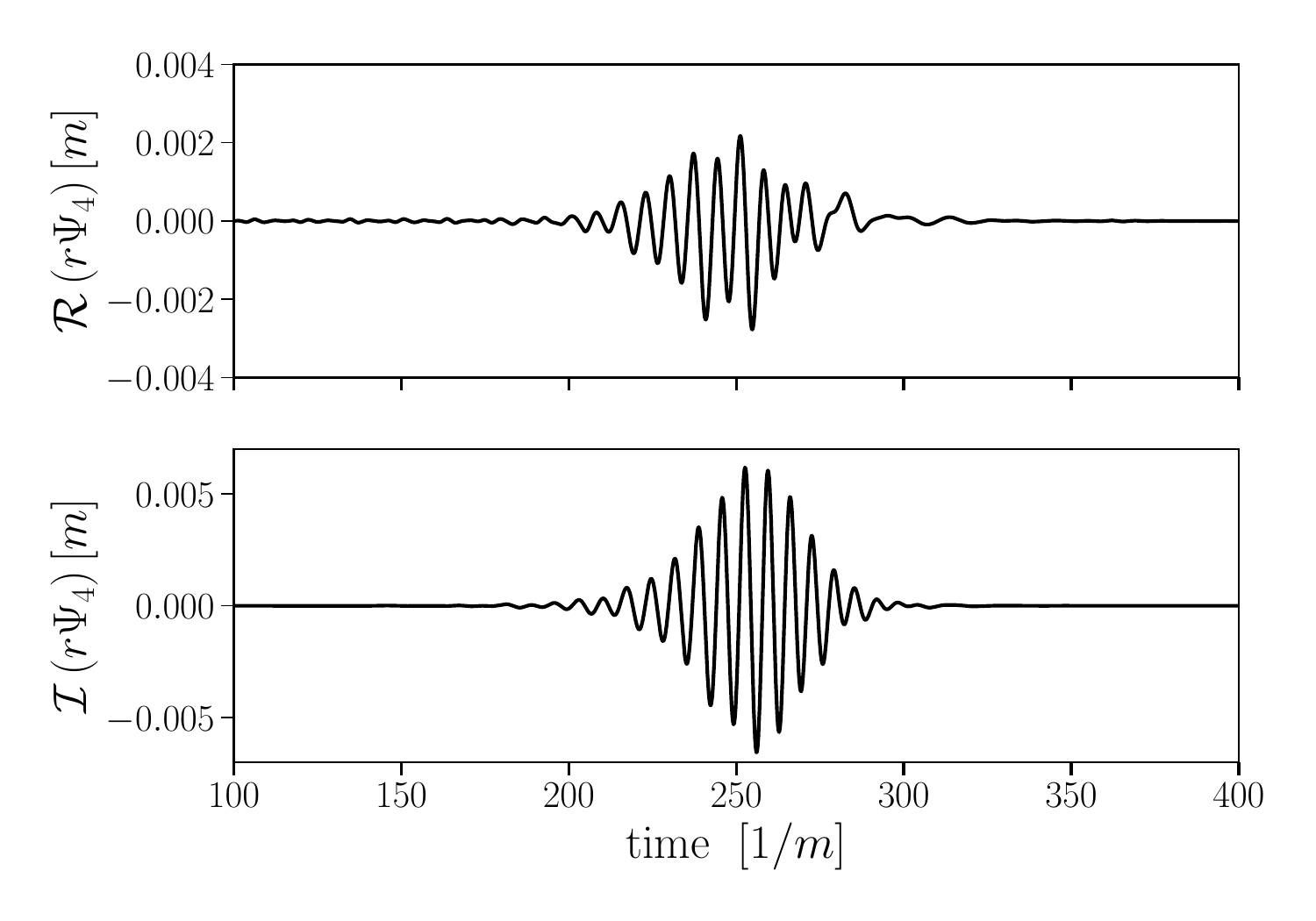}
        \vspace{-0.7cm}
    \caption{Gravitational wave waveform of the $l=2, m=0$ mode for the electromagnetic decay of a single star. The waveform is extracted at $r=80m^{-1}$ at refinement level 2. This gravitational wave, corresponding to a burst of electromagnetic energy, does not exhibit a ringdown. We note that the $l=2, m=\pm2$ modes were of similar order of magnitude.}
    \label{fig: GW-single}
        \vspace{-0.5cm}
\end{figure}

In Fig.~\ref{fig: GW-single} we show the gravitational wave signal corresponding to the electromagnetic decay of a single compact star with mass $M_s = 0.60 m_{\mathrm{Pl}}^2/m$, using $g_{a\gamma} = 16 m_\mathrm{Pl}^{-1}$. The two polarisation modes of GWs are captured through the real and imaginary parts of the Newton-Penrose Weyl scalars $\Psi_4$ (see, e.g. Ref.\cite{bands}). Fig.~\ref{fig: GW-single} shows the real and imaginary part of the $l=2, m=0$ mode, although the $l=2, m = \pm2$ modes were observed to be of similar amplitudes. The signal is extracted at $r=80m^{-1}$. 

\begin{figure*}[]
    \centering
    \includegraphics[width=13.0cm]{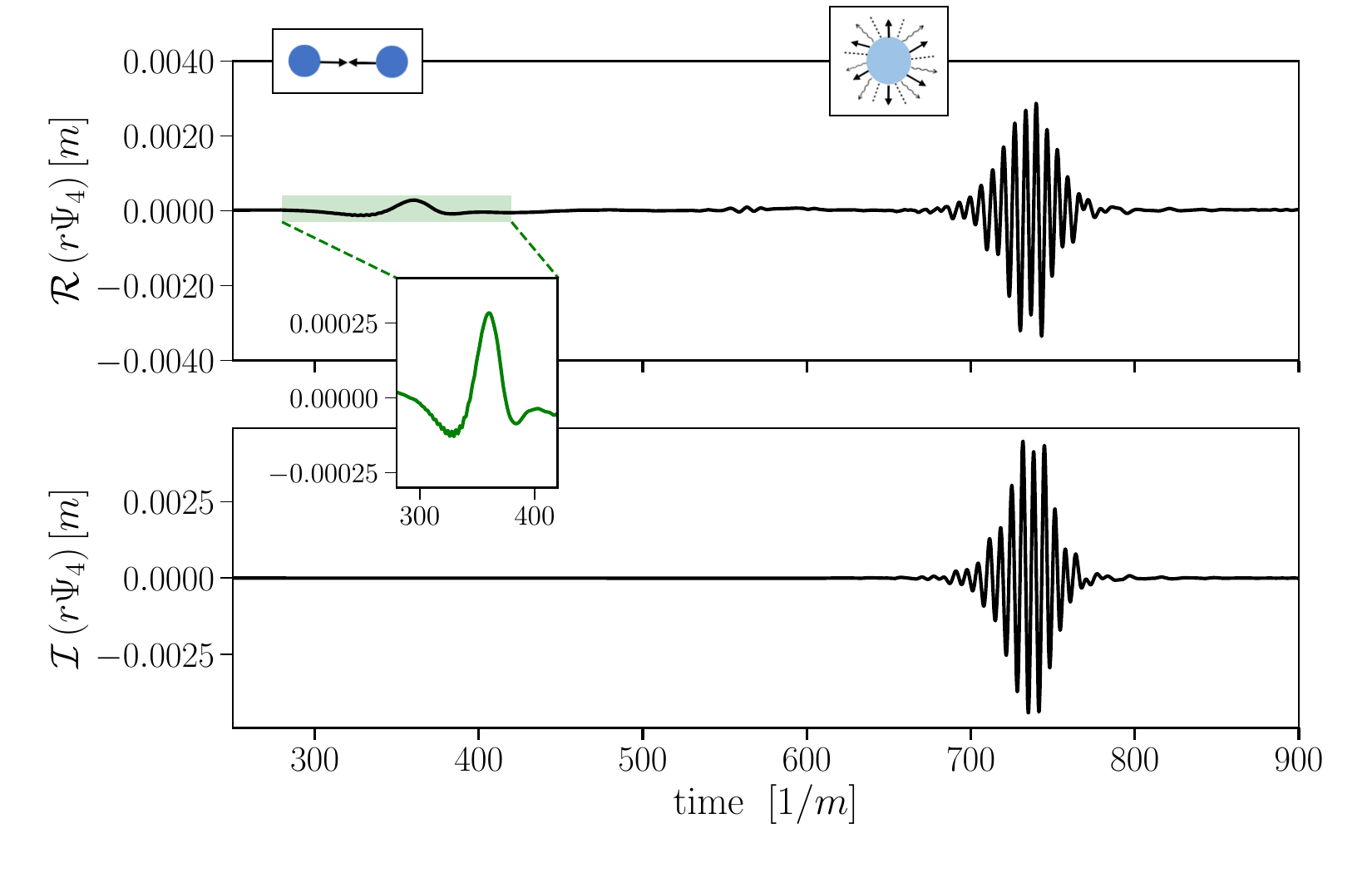}
        \vspace{-0.5cm}
    \caption{Gravitational wave waveforms for the merger and subsequent decay process, showing the dominant quadrupole mode $l=2, m=0$. We note that the $l=2, m = \pm2$ modes were of the same order of magnitude than the mode shown. The wavefronts were extracted at radius $r=80m^{-1}$, refinement level 2. The first excitation around $t=350 m^{-1}$ in the real part corresponds to the merger signal, and the second excitation around $t=750 m^{-1}$ is the signal from the electromagnetic decay of the star. In the case of the decay, both the real and imaginary parts are excited, meaning that both gravitational polarisation modes are present. A zoomed in picture of the signal from the merger of the stars is depicted as a green wavefront, corresponding to the shaded green region in the main plot.}
\label{fig:GW-full}
    \vspace{-0.5cm}
\end{figure*}

As the decay of the star is not symmetric (for example, our electromagnetic seed is polarised in one direction and propagating in another), both gravitational polarisation modes are excited, corresponding to the real and imaginary parts of the Weyl scalar. In contrast, the merger signal is axisymmetric, and hence only a single polarisation mode is excited. We also notice that this decay signal does not have a ringdown hierarchy like black hole mergers, creating a distinct signature. 

Moreover, in Fig. \ref{fig:GW-full} we map the whole gravitational wave waveform for the merger of two stars and the subsequent parametric decay process, outlined in Section \ref{sec:mergerstory}. The gravitational wave signal from the merger at around $t \sim 350 m^{-1}$ has already been studied in \cite{Helfer:2018vtq}. We note that the GW signal from the decay of the merged star into photons around $t \sim 750 m^{-1}$ closely matches the signal from the decay of a single star shown in Fig.~\ref{fig: GW-single}.

The frequency of the oscillations in the GW signal were computed by taking a Fourier transform, and found to be around $\omega \sim m$ for the decay, which matches the frequency of the EM flux through the same surface (see Section \ref{sec:multimess}), and moreover the mass of the scalar field. The frequency match $\omega \sim m$ can be understood from linearising the axion-GW interaction. The frequency of the merger signal, on the other hand, is around one order of magnitude lower, $\omega \sim 0.1m$. For an axion mass of $m=10^{-12}\text{ eV}$, the decay frequency corresponds to $f \sim 10^3\mathrm{Hz}$, implying that both the decay and the merger frequencies lie close to the current peak sensitivity of the LIGO band.

\begin{figure}[b!]
    \centering
    \includegraphics[width=9cm]{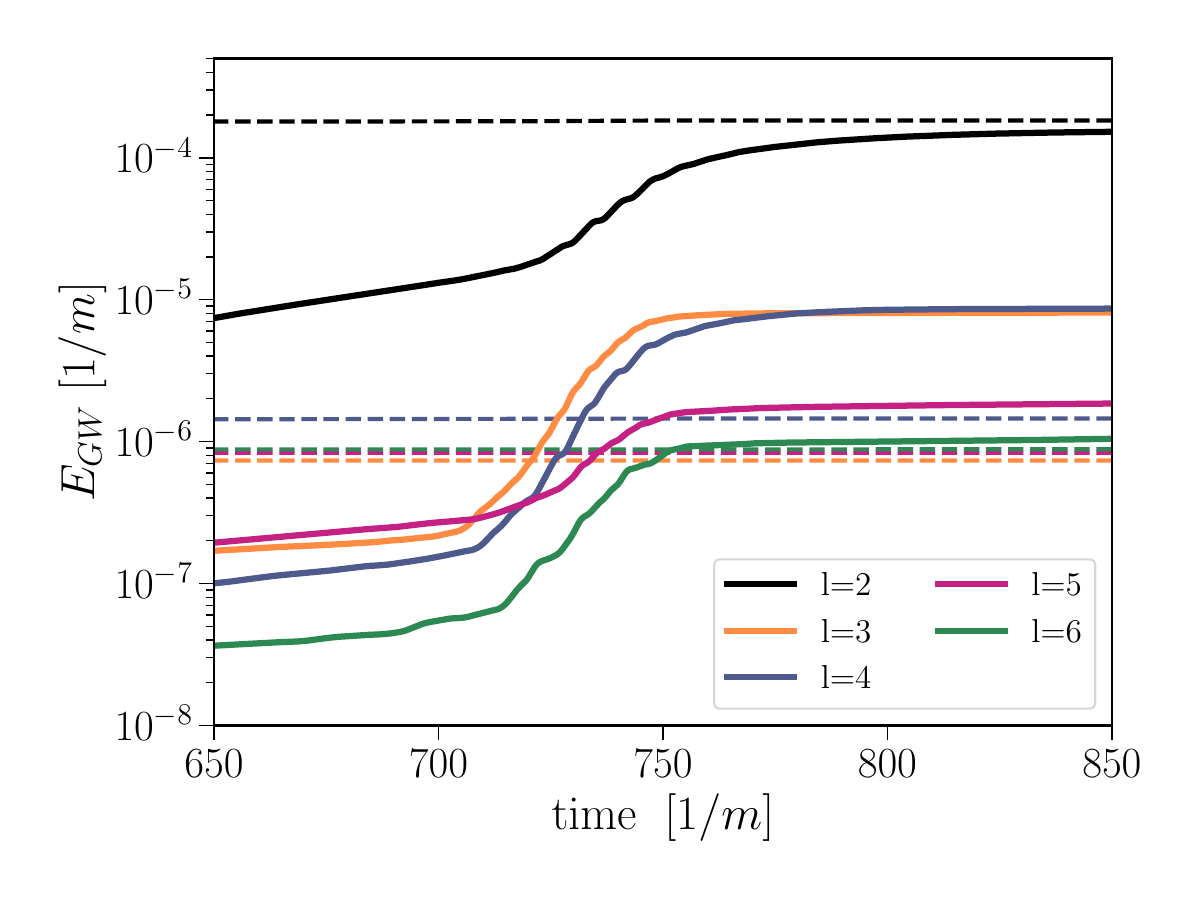}
        \vspace{-0.7cm}
    \caption{Cumulative gravitational energy in each $l$ mode for the decay of the merged star (corresponding to the signal at $t \sim 750m^{-1}$ in Fig.~\ref{fig:GW-full}). The dashed lines indicate the maximum energy of the corresponding mode for the single star decay. We note that the $l=2$ mode dominates, as expected, and that the spectrum of the merged star is more varied than the single one, where most of the energy is in the dominant $l=2$ mode.}
    \label{fig: GW-energy}
        \vspace{-0.5cm}
\end{figure}

Finally, we study the gravitational wave energy resulting from the electromagnetic decay of the star. Fig.~\ref{fig: GW-energy} shows the cumulative energy in the gravitational wave signal from the decay of the merged star for the dominant $l$ modes, extracted from the signal between $t=600m^{-1}$ to $t=900m^{-1}$. This spans the decay process of the merged star. As expected from other compact objects, the dominant mode is the quadrupole, $l=2$. Comparing these modes to the modes excited by the decay of the single star (maximum energy in each mode shown as dashed lines), we note that the spectrum of the merged star has more power in higher multipoles as the merger excites asymmetric spatially oscillating modes in the star (the single star, on the other hand, is spherically symmetric before the decay process). Concretely, less than 1\% of the GW energy from the decay of a single star is found in modes $l>2$, whereas the merged star has around 10\% of the energy in the $l=3$ and $l=4$ modes.

The total gravitational energy ($E_{GW}$) from the decay is $\sim 0.03\%$ of the total energy of the original stars ($2M_s$), comparable to the ratio in black hole head-on mergers. We expect this efficiency to be similar for non-head on mergers as it is driven by the decay process. On the other hand, the electromagnetic decay of the merged axion star carries around ten times more energy in gravitational waves than the merger of axion stars. 

Our simulations also demonstrate that the total gravitational energy emitted by the single star decay scales roughly linearly with compactness, with an increase of over $90\%$ in the energy emitted from $C=0.10$ to $C=0.14$, keeping the coupling constant fixed at $g_{a\gamma} = 16 m_\mathrm{Pl}^{-1}$. However, as the value of the axion-photon coupling affects the decay process significantly, a multi-parameter study would be required for a full understanding of the energy scaling.

\section{Multimessenger signal} \label{sec:multimess}

\begin{figure*}[]
    \centering
    \includegraphics[width=19.0cm]{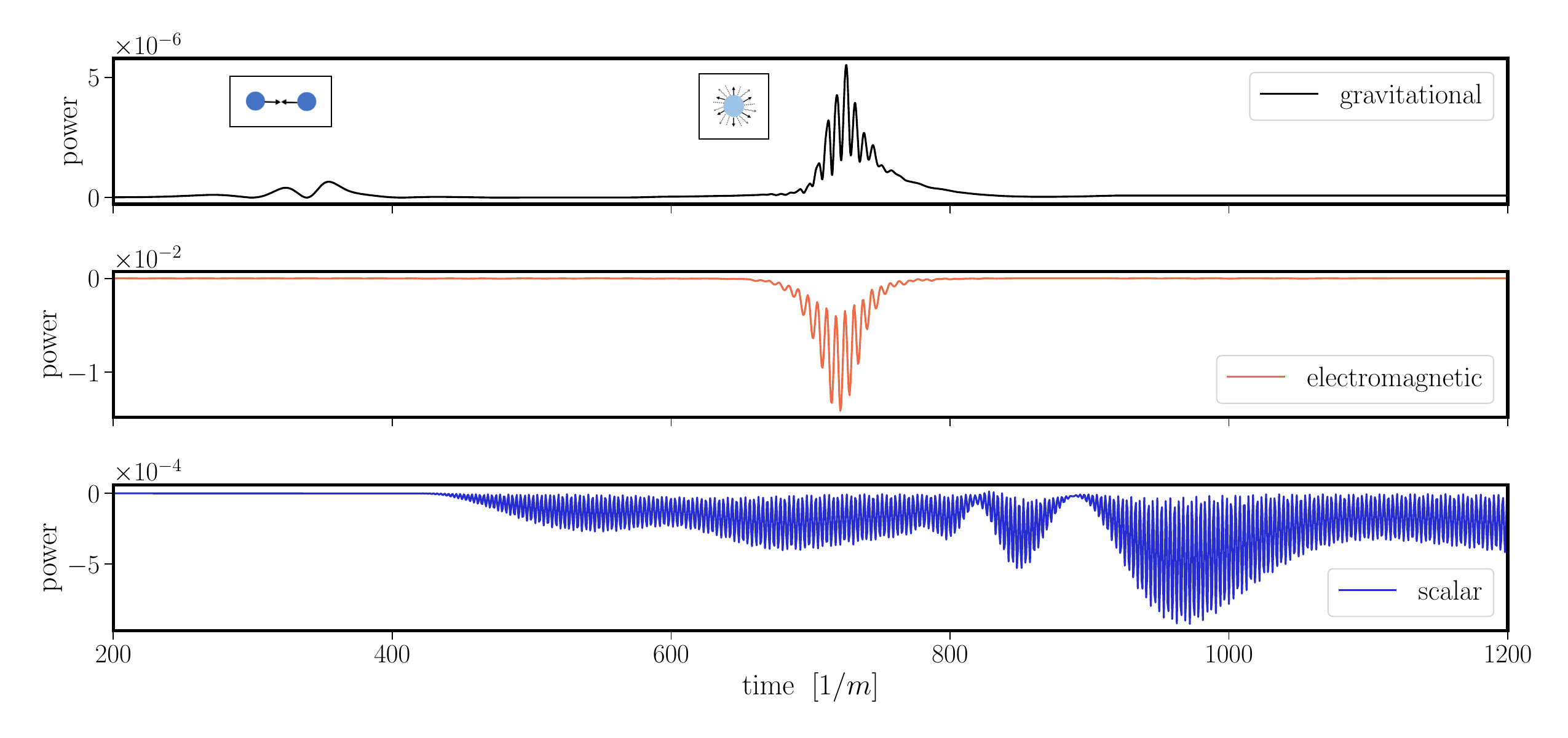}
        \vspace{-0.5cm}
    \caption{Multimessenger signal from the merger of two axion stars and subsequent decay of the merged star, caused by an electromagnetic instability: the gravitational power (\textit{top}), total flux of electromagnetic energy extracted over a sphere at the same radius as the GW signal, $r=80m^{-1}$ (\textit{middle}), and corresponding total flux of the scalar energy (\textit{bottom}), where the power in electromagnetic and scalar radiation is obtained by integrating the corresponding flux. The different frequency wave modes in the scalar emission are responsible for the long tail in the scalar flux, as lower frequency modes travel slower away from the decaying star. The system starts shedding scalar waves already due to the merger which happens around $t \sim 300 m^{-1}$, resulting in a complex pattern of scalar radiation after the electromagnetic decay (around $t \sim 750 m^{-1}$).}
\label{fig:multimessenger_signal}
    \vspace{-0.5cm}
\end{figure*}

As the previous sections have explored in detail, the electromagnetic instability of the axion star results in a multimessenger signal, with both an electromagnetic and a gravitational component. The gravitational wavefront and the electromagnetic energy ejected are expected to reach an observer away from the source at the same time, travelling at the speed of light. This is represented in Fig.~\ref{fig:multimessenger_signal_single}, where the upper two panels represent the gravitational power (in black) from the decay of a single star (corresponding to the wavefront in Fig.~\ref{fig: GW-single}), and the corresponding total flux of electromagnetic energy (in orange), integrated over a sphere at the same radius ($r=80m^{-1}$).

The bottom panel (in blue) represents the scalar (axion) waves ejected by the star, captured by the total scalar flux over the same sphere. These total energy fluxes are calculated by integrating over the flux in Eq. 11 of Ref. \cite{Clough:2021qlv}. Firstly, we observe that the scalar energy is around two orders of magnitude lower than its electromagnetic counterpart (the power units are the same in both panels), hence being a subdominant effect. Secondly, the total scalar flux is delayed compared to the EM and GW signals, as massive modes travel slower than light~\footnote{To confirm the delay of the massive modes compared to the EM and GW signal, we computed the expected time delay for the single star decay using the relativistic dispersion relation and $k \sim 0.5m$, which corresponds to the characteristic frequency of the star, assuming it matches the maximum group velocity. This gives us $\Delta t \sim 100 m^{-1}$, which matches the time delay in Fig.\ref{fig:multimessenger_signal_single}.}. In addition, in the decay process, different wave modes of the scalar field are excited as the star transitions from one state to another. These scalar waves of different wavelengths travel at different speeds, resulting in a long tail of total scalar flux as shown in Fig.~\ref{fig:multimessenger_signal_single}, as longer wavelength modes travel slower (recall that $\omega^2=k^2+m^2$ for a free scalar wave) resulting in the characteristic ``dispersion'' of the wave packet.

In Fig.~\ref{fig:multimessenger_signal} we show the corresponding GW, EM and scalar signals for the merger simulations. The GW and EM signals from the decay phase are strikingly similar to the single star case, with a relatively small GW bump from the merger. On the other hand, the merged star emits scalar waves immediately following the merger at $t \sim 300m^{-1}$, due to being in an excited state \cite{Seidel:1993zk} ending in a burst during the electromagnetic decay phase.

Finally, we comment on the energy scales of the system: the largest source of emitted energy in the parametric, electromagnetic decay of an axion star is the electromagnetic emission from the star, which is around two orders of magnitude larger than the scalar emission. The gravitational energy emitted is a further two orders of magnitude smaller than the scalar emission, hence creating a clear hierarchy in the magnitude of the signals emitted. For a scalar field mass of $m=10^{-12}\text{ eV}$, the total electromagnetic energy ejected by the parametric decay of the merged star is around $10^{49}\mathrm{J}$, which compares to the energy scale of a supernova.

\section{Discussion and Conclusions} \label{sec:discuss}

In this work we have studied, in fully non-linear simulations, the multimessenger signal produced by the collision of two compact axion stars close to the critical mass defined by the axion-photon coupling. The initial merger produces a GW signal consistent with that expected for real scalar boson stars~\cite{Giudice:2016zpa, Helfer:2018vtq}. This signal is relatively small in amplitude, due to the low compactness of $C<0.14$. The merger forms an excited, more massive axion star that crosses the threshold of electromagnetic stability. This star subsequently decays into electromagnetic radiation via parametric resonance, triggered by ambient photons. The excited super-critical star first cools by emission of axions, and then explodes, emitting axions, GWs, and photons. The GWs from the decay are almost monochromatic, at a frequency one order of magnitude larger than the GWs from collision, and do not follow a ringdown hierarchy. GWs from the decay carry over one order of magnitude more energy than those from the merger. 

The length of time of the delay between merger and decay is fixed by the coupling $g_{a\gamma}$, the mass of the merged star (which defines its critical coupling), as well as the amplitude of the EM seed through a logarithmic dependence, as we confirmed in \cite{Chung-Jukko:2023cow}. The dominant energy in the decay is carried by photons, since criticality is governed by parametric resonance of the Chern-Simons interaction between axions and photons. The energy in photons is two orders of magnitude larger than the energy in axions, which itself is two orders of magnitude more than the energy in GWs. 

This paper has been concerned with the theory and simulation of axion star mergers and explosions, computing for the first time the decay channels. We discuss briefly here the observational consequences, but leave full study to a future work. By colliding two axion stars of mass $M_s = 0.36 m_{\mathrm{Pl}}^2/m$, $C=0.03$ each, with axion parameters $m=10^{-12}\text{ eV}, g_{a\gamma} = 20 m_\mathrm{Pl}^{-1}$, selected to give rise to GWs from merger and decay in the LIGO frequency band, we determined that the GWs would be visible up to distances of $\sim 110 \text{ Mpc}$ and $270 \text{ Mpc}$, for merger and decay respectively. The calculations follow Eq. 13 in \cite{Helfer:2018qgv}, assuming a detected strain $h = 10^{-22}$, which is the current LIGO peak sensitivity for the decay frequency range. If GW emission and photon emission from the decay could both be observed, then with the right waveform model and timing data this may allow the weaker merger GW signal to be extracted from noise in post-processing.

An axion star explosion from higher mass axions produces high frequency GWs, for example $m\approx 10^{-6}\text{ eV}$ gives GWs in the GHz band, although the luminosity would be smaller as the axion star mass scales inversely to $m$. Such high frequency GWs can be detected using microwave cavities~\cite{Berlin:2021txa} (such as the ones commonly used in axion dark matter haloscopes) or phononic materials~\cite{Kahn:2023mrj} . We have not considered this possibility in detail for GWs from axion star explosions, as the signal is likely too weak to give sensitivity to large astronomical distances, but this may warrant further investigation.

For GWs in the LIGO band, the EM radiation is emitted in the kHz band with total energy comparable to a supernova. Some consequences of the EM radiation emitted by an axion star explosion in this band are discussed in Refs.~\cite{Escudero:2023vgv,Du:2023jxh}. However, for a multimessenger signal including GWs, there are several important differences to consider for the photon emission compared to Refs.~\cite{Escudero:2023vgv,Du:2023jxh}. Compact stars giving a GW signal require a different formation mechanism: the core halo-mass relation~\cite{Schive:2014dra} for axion stars in dark matter halos predicts that compact axion stars are extremely rare, if present at all. One possible mechanism to form compact axion stars is outlined in Ref.~\cite{Widdicombe:2018oeo}, which involves a spike in the axion primordial power spectrum, which could emerge due to phase transitions and string network decay after inflation (e.g. as axiton formation in Ref.~\cite{Vaquero:2018tib}), or blue inflationary isocurvature~\cite{Kasuya:2009up}. In such mechanisms, compact axion stars can form at rare density peaks, much like primordial black holes. Another possibility involves an extended matter dominated epoch~\cite{Erickcek:2011us}. The physics of an individual explosion will be the same as studied in Refs.~\cite{Escudero:2023vgv,Du:2023jxh}, but the population level effects will be absent.  Furthermore, a multimessenger signal involving GWs will probe only the local Universe at $D\lesssim 300\text{Mpc}$, rather than the early universe $z>10$ considered in Refs.~\cite{Escudero:2023vgv,Du:2023jxh}.

The EM emission from the explosion of an axion star in the kHz band is efficiently absorbed by neutral hydrogen, causing it to ionize~\cite{Bolliet:2020ofj}. Thus, we would observe the EM emission indirectly from the heating of the surrounding gas. Following Ref.~\cite{Escudero:2023vgv,Du:2023jxh} we find that in the Milky Way, taking values for the free electron density and baryon temperature in virial equilibrium, the absorption length for photons in the kHz band is short, on the order of parsecs. Thus the photons from an exploding axion star in the local Universe heat the surrounding gas and rapidly ionize and shock it;  the resulting emission in optical frequencies from the hot expanding gas may resemble a supernova with no associated production of heavy elements.

It may also be possible to detect the semi-relativistic axions produced. Axion dark matter haloscope data is normally analysed assuming a search for the cold galactic axions of the standard halo model~\cite{Brubaker:2017rna}. However, the physics of axion-photon conversion can happen also for relativistic axions, such as those of the ``cosmic axion background''~\cite{Dror:2021nyr}. In particular, the broadband DMRadio proposal~\cite{DMRadio:2022jfv} covers the correct frequency range for axions that are part of a multimessenger signal of an axion star explosion with accompanying GWs in the LIGO band.

\section*{Acknowledgements} 

We would like to thank Lewis Croney, Miguel Escudero, Tamara Evstafyeva and Thomas Helfer for useful discussions. We also thank members of the \textsc{GRTL} Collaboration for technical support and help. LCJ is supported by a studentship funded by the Science and Technologies Facilities Council (UK). DJEM is supported by an Ernest Rutherford Fellowship from the Science and Technologies Facilities Council (ST/T004037/1). This work used the DiRAC@Durham Cosma facility managed by the Institute for Computational Cosmology on behalf of the STFC DiRAC HPC Facility (www.dirac.ac.uk), under DiRAC grant ACTP238. The equipment was funded by BEIS capital funding via STFC capital grants ST/P002293/1, ST/R002371/1 and ST/S002502/1, Durham University and STFC operations grant ST/R000832/1.

\bibliography{bosonstar}

\clearpage

\appendix

\section{Shutting down of resonance} \label{sec:resshut}

\begin{figure}[h!]
\centering
\includegraphics[width=\linewidth]{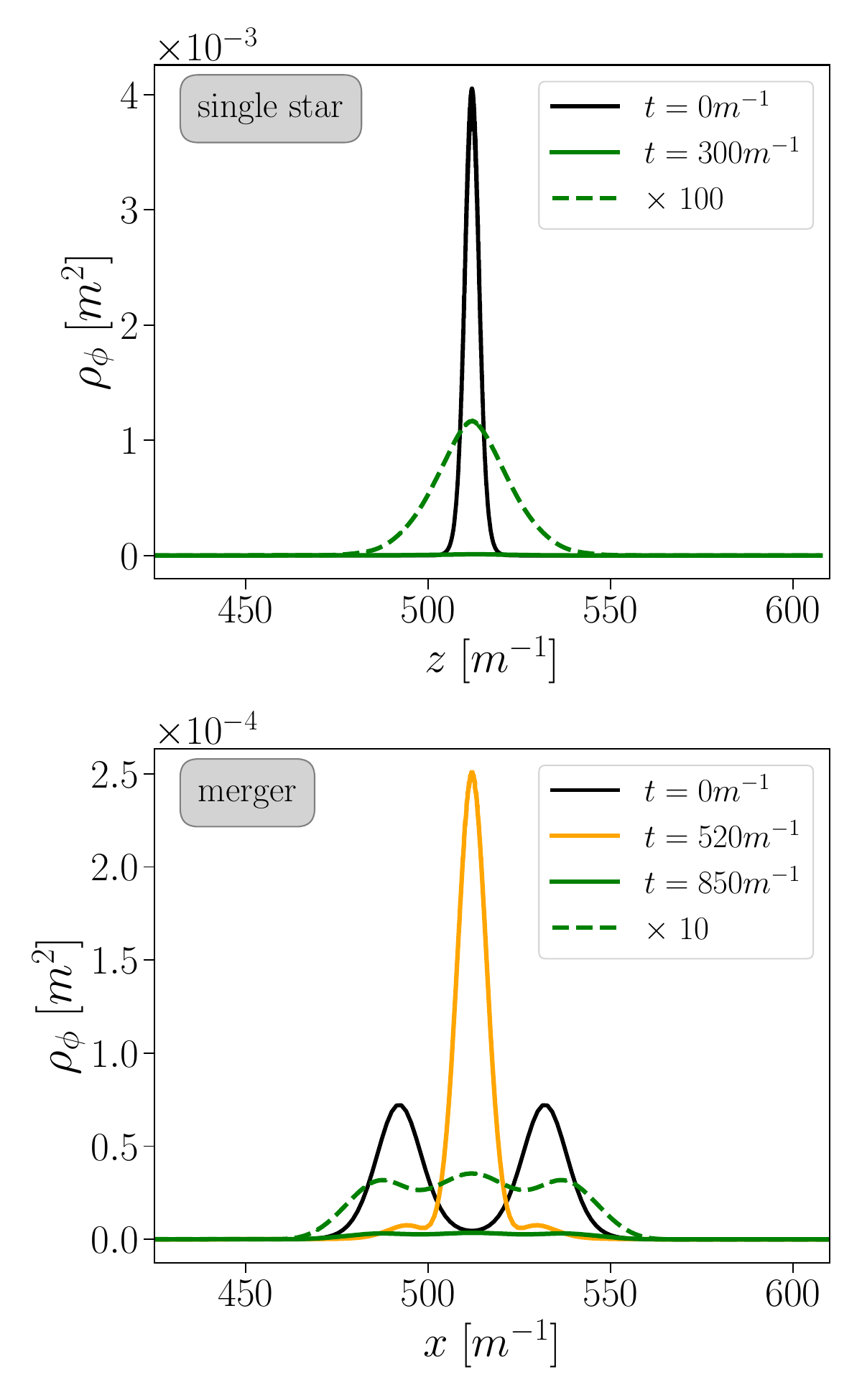}
\caption{Radial profiles of the scalar energy density $\rho_{\phi}$ of a single star decay (\textit{top}) and the merger process (\textit{bottom}). The original set up is shown in black (single star on the top and two stars on the bottom plot), and the remnant after the electromagnetic decay has happened, in green. The dashed green line is a magnification of the remnant energy density, to demonstrate its energy distribution, which indeed has a larger radius and broader shape than the corresponding original star, as the star has diluted in the parametric resonance process. The magnification factors are 100 and 10, for the single and merged stars, respectively. The merged star created trough the collision of the two sigle stars in the bottom plot, which subsequently decays through parametric resonance, is shown in orange. We note that these radial profiles have been taken through the center of the simulation box, with the other two coordinates at $512m^{-1}$, and we have chosen the $x$-axis for the merger process, as the stars are originally aligned along $x$.}
\label{fig:radprof}
\end{figure}

Axion stars can experience an electromagnetic instability if their characteristic frequency lies within the electromagnetic resonance band, as we first demonstrated in \cite{Chung-Jukko:2023cow}, and further in this work for the case of a merged star. In Fig. \ref{fig:radprof}, we show how the parametric resonance process responsible for the decay of the star into electromagnetic radiation does not dissipate the whole star, but shuts down due to the dilution of the star. 

As the star enters parametric resonance, its shedding of energy causes the star to dilute, increasing its size. Since the characteristic frequency of the star is linked to its size, the change in frequency means that the star slides off the resonance frequency band and hence the process stops, leaving behind a less compact, dilute remnant. This is seen in the upper plot of Fig. \ref{fig:radprof} for the single star, where the original star represented by the black line has decayed into a significantly less energetic, less compact star, shown in green. The magnification of the energy density of the remnant star, shown with a dashed green line, demonstrates its broader energy distribution. This indicates a larger radius, and hence different characteristic frequency to the original star.

Similarly, in the bottom plot the original two stars are shown in black, the merged star created by their collision in orange, and the remnant after the electromagnetic decay in green, with the magnification of the remnant energy with a dashed green line. The corresponding times, shown in the legend, can be mapped to the process represented as a whole in Fig. \ref{fig: collenergy}. As the merged star is created by the collision of the single stars, its energy distribution has some additional features compared to the single star, which are discussed in Section \ref{sec:mergerstory}. These features are also present in the remnant, which exhibits a broader energy distribution and hence a larger radius than the original stars, as expected.

\section{Convergence testing}  \label{sec:convergencetesting}

\subsection{General convergence}

\begin{figure}[t!]
\centering
\includegraphics[width=\linewidth]{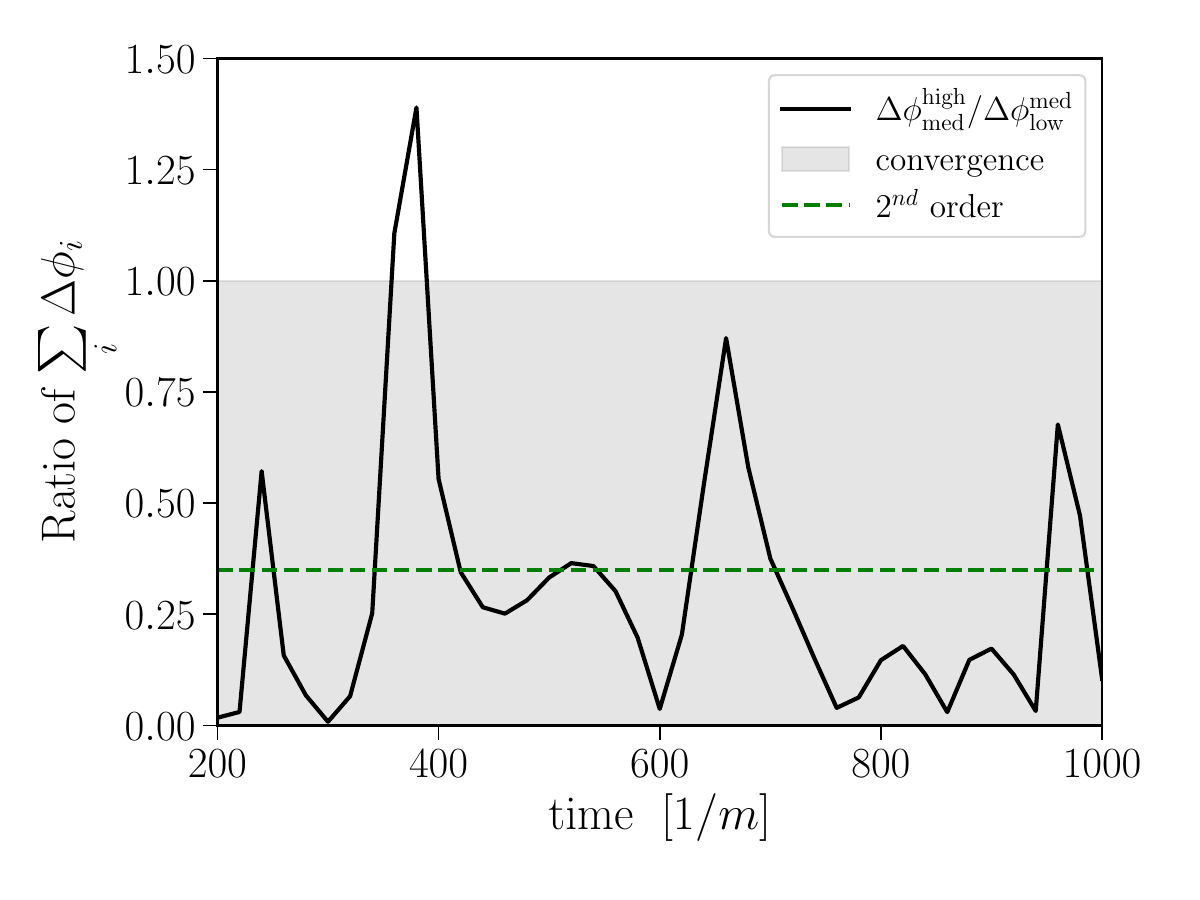}
\caption{Convergence testing using a stencil of 7 points centered in the middle of the simulation box. We plot the ratio of the difference in the scalar field $\phi$ value between the high and medium, and the medium and low resolution runs, summing over the stencil. Hence, all data points within the grey region indicate convergence. The simulation box size is $L=1024m^{-1}$, and low, medium, and high resolution runs have $N^3=256^3$, $N^3=384^3$, and $N^3=512^3$ number of coarse grid points, respectively.}
\label{fig:convergence}
\end{figure}

We demonstrate the convergence of our simulation by monitoring the ratio of the difference in the scalar field $\phi$ value between the high and medium, and the medium and low resolution runs, summing over a stencil of 7 points, centered at the middle of our simulation box. The box had a size $L=1024m^{-1}$ and a maximum of 7 refinement levels. The number of coarse grid points used for each resolution was $N^3=256^3$, $N^3=384^3$, and $N^3=512^3$ (from lower to higher resolution). The region of convergence, defined by the ratio being below 1, is demonstrated by the grey area in Fig.~\ref{fig:convergence}, alongside the ratio values (shown with the black solid line), as a function of time. 

We note that there is a peak at around $t \sim 400 m^{-1}$ which does not fall in the region of convergence. This is due to the regridding forced on the simulation by the gravitational wave extraction tool. Our numerical relativity code \textsc{GRChombo} includes an adaptive mesh refinement system that automatically adds refinement levels in specific regions of the simulation box, based on set conditions and threshold values. However, the extraction of gravitational waves at a set radius and refinement level overrides the original, smooth refinement process by forcing a specific refinement layer on the simulation at each time step. As the boundaries of each layer introduce numerical errors to the simulation, we postulate that it is this regridding process that is responsible for the peak in our convergence plot. This was shown by running the same simulation without GW extraction, in which case the peak at around $t \sim 400 m^{-1}$ disappeared. Moreover, the differences in the $\phi$ value as shown in Fig.~\ref{fig:convergence} get smaller as the simulation progresses, which is consistent with convergence. 

\subsection{Gravitational wave convergence}

\begin{figure}[h!]
\centering
\includegraphics[width=9.5cm]{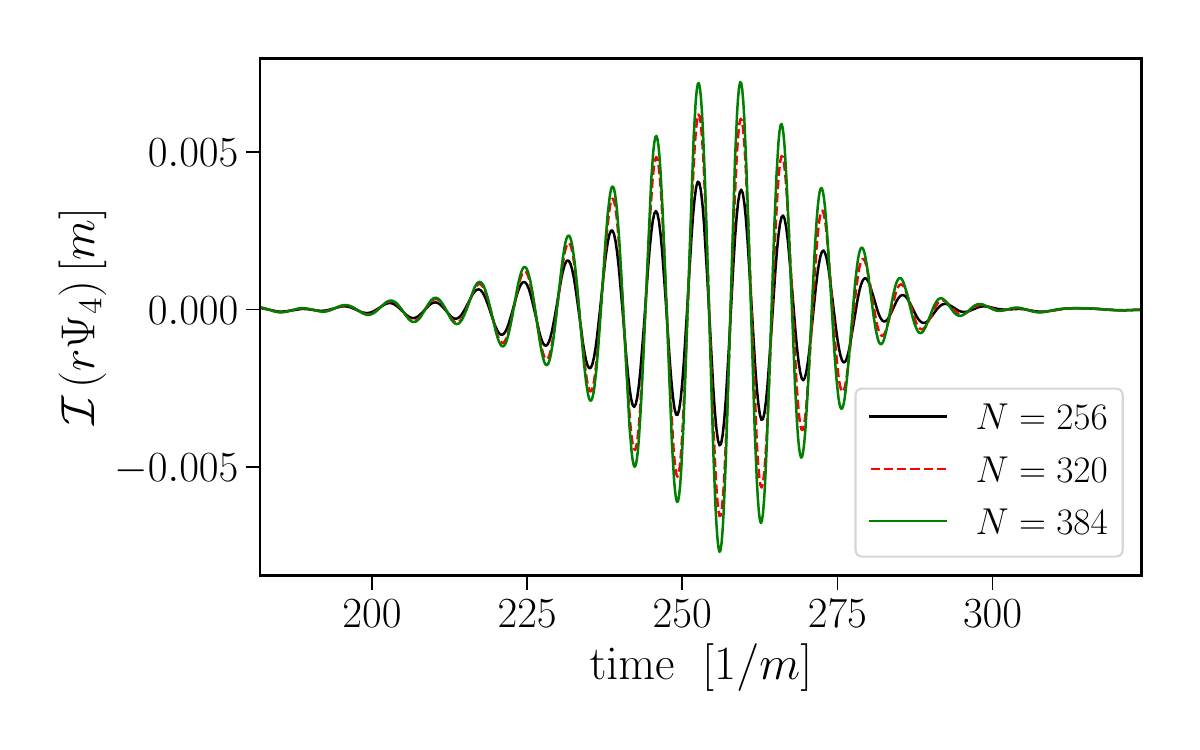}
\caption{Convergence testing of the gravitational wave waveform for the decay of a single star. The number of coarse grid points used for each resolution is shown in the legend. Convergence can be seen by eye as the higher resolutions (in green and red dashed) are closer together than the lower ones (red dashed and black). The waveforms were extracted at refinement level 2 at $r=80 m^{-1}$.}
\label{fig:GWconvergence}
\end{figure}

In addition to testing the convergence of the bulk simulation, as explained above, we monitored the convergence of the gravitational wave waveforms, which is demonstrated in Fig.~\ref{fig:GWconvergence}. The figure shows the imaginary part of the $l=2, m=0$ mode of the Weyl scalar, but we note that the real part followed a similar pattern. Convergence can be clearly seen by eye, as the difference between the high and medium resolution waveforms is smaller than that between the medium and the low resolutions. 

Here we demonstrate the convergence for the GW signal from the decay of a single, stable star of mass $M_s = 0.60 m_{\mathrm{Pl}}^2/m$. The merger GW signal (shown in Fig.~\ref{fig:GW-full}) had better convergence. While Fig.~\ref{fig:GWconvergence} shows the signal from the decay of a single (non-excited) star, we also confirmed the convergence for the full combined merger and decay process. Due to the high frequency of the gravitational wave waveform of the decay, it was challenging to keep track of it, and indeed we found linear drift in the waveforms between different resolution runs for the decay after merger, corresponding to 3-4 wavelengths in time. However, the frequencies of the different resolution waveforms matched. Overall, despite the simulation and gravitational waveforms converging, we found some variation in the waveform amplitudes.

\subsection{Regridding}

In order to extract an overall clean waveform, we had to tune our regridding for the merger and the decay signal separately, using our adaptive mesh refinement code. The goal of regridding is to add higher resolution meshes in the simulation box areas where the interesting physics, and hence the need for higher resolution, is happening. For our simulations, this means that the star is resolved with a higher resolution grid than the empty space around it. (However, as mentioned above, the extraction of gravitational waves away from the star, closer to the boundaries of the box, means we also force a higher resolution at the extraction radius). In practice, this tuning of the regridding means that we had to change the threshold values for the variables with respect to which the refinement is happening (in our case, the scalar field $\phi$ and the conformal factor $\chi$). 

To get the full signal as shown in Fig.~\ref{fig:GW-full}, we had to decrease the thresholds between the merger and the decay signal. The merger signal exhibited some numerical noise that was reduced with lesser regridding, while the high oscillation frequency of the decay signal required more regridding and higher extraction levels. The decrease in the regridding thresholds (hence, more frequent regridding), moving from the merger to the decay, was $18\%$ and $23\%$, for $\phi$ and $\chi$, respectively. This regridding change had no effect on the overall evolution of the system, and demonstrates the difficulty of gravitational wave extraction from our oscillating scalar star. 

Finally, we note that we found some linear drift in time in the energy fluxes and gravitational wave signals from the simulation due to this change of regridding in the middle of the run (between the merger and the decay of the merged star). Hence, the gravitational signal in Fig. ~\ref{fig:multimessenger_signal} has been corrected to match the electromagnetic one, as we confirmed by running the simulation from the middle that those two indeed line up (our full GW signal displayed in Fig. ~\ref{fig:GW-full} has the regridding change in the middle, as explained above).  

\section{Gravitational luminosity}  \label{sec:numerics}

The gravitational wave power (luminosity) was calculated according to the well-known equation (see e.g. \cite{Croft:2022bxq} for similar use):

\begin{equation} \label{eq:GWlum}
\frac{d E}{d t}=\lim _{r \rightarrow \infty} \frac{1}{16 \pi} \sum_{l, m} \left|\int_{t_0}^t r \Psi_{4}^{lm} d t^{\prime}\right|^2,
\end{equation}

where we used the modes up to $l=6$ for the decay signal of the star and $l=2$ for the merger signal of the star. To deal with the numerical error causing a cumulative drift in the total gravitational energy of the system, the integral of $r \Psi_{4}^{lm}$ was processed with a Butterworth high-pass filter (5th order), as the numerical error introduces low-frequency modes.

\clearpage

\end{document}